\documentstyle[prl,aps,amsmath,amssymb,amsthm]{revtex}

\begin{document}
\pagenumbering{arabic}
\setcounter{page}{1}

\title{Lower bound on the relative error of mixed-state cloning and
related operations}

\author{ A. E. Rastegin }
\address{Department of Theoretical Physics, Irkutsk State University,
Gagarin Bv. 20, Irkutsk 664003, Russia \\
Electronic Mail: {\tt rast@api.isu.runnet.ru}}

\maketitle
\begin{abstract}
We extend the concept of the relative error to mixed-state
cloning and related physical operations, in which the ancilla
contains some {\it a priori} information about the input state. The
lower bound on the relative error is obtained. It is shown that this
result contributes to the stronger no-cloning theorem.

\end{abstract}

\protect\section{Introduction}

Quantum cloning is a important issue in quantum information, due to
its connection to security in quantum cryptography and reflection on
the nature of quantum states. It is well known that nonorthogonal
pure states cannot be cloned \cite{wootters}. This result
was generalized and extended in Ref. \cite{barnum}: noncommuting
mixed states cannot be broadcast. In Ref. \cite{jozsa1} the stronger
no-cloning theorem was established. For example, let
$\,\{|s_1\rangle,|s_2\rangle\}\,$ be any pair of nonorthogonal
pure states and $\,\{\Upsilon_1,\Upsilon_2\}\,$ be any pair of mixed
states. According to the stronger no-cloning theorem, there
is a physical operation
$\:|s_j\rangle\otimes\Upsilon_j \longmapsto |s_j\rangle
|s_j\rangle\:$
if and only if there is a physical operation
$\:\Upsilon_j \longmapsto |s_j\rangle\:$. In other words, the full
information of the clone must be {\it a priori} provided in the
ancilla state $\Upsilon_j$ alone \cite{jozsa1}.

The approximate quantum copying was originally considered by
Bu\v{z}ek and Hillery\cite{buzek}. In addition, they examined
approximate cloning machines operating on prescribed two
non-orthogonal states \cite{hillery}. In Ref. \cite{brass} such
devices were called 'state-dependent cloners'. As a criterion for
estimation of the state-dependent cloning, Ref. \cite{brass}
introduced "global fidelity" and "local fidelity." It has constructed
the optimal "global" cloner that maximizes the global fidelity. The
local fidelity has also been optimized. The writers of Ref.
\cite{chefles} obtained the upper bound on the global fidelity for
$N\to L$ cloning of two states with {\it a priori} probabilities.
Ref. \cite{macchi} considered state-dependent $N\to L$ cloning with
respect to both the mentioned criteria.

The other category of cloners contains universal cloning machines
which copy arbitrary state equally well. First such example was given
by Bu\v{z}ek and Hillery \cite{buzek}. Refs. \cite{brass,gisin}
constructed the universal qubit cloner that maximizes the
local fidelity. Analogous problem for multi-level quantum system was
solved in Refs. \cite{werner,keyl}. Note that the approximate cloning
is interesting for several questions. The problem of security in
quantum criptography is one of obvious applications. In addition, the
cloning transformations can be used to realize joint measurements of
noncommuting observables \cite{macchi1}.

Thus, the state-dependent cloning was mainly examined from the
"fidelity" viewpoint. However, the state-dependent cloning is
a complex subject with many facets. Important as the notions of the
global fidelity and the local fidelity are, they do not cover the
problem on the whole \cite{rastegin1}. An optimality criterion to
widen an outlook is needed. In Ref. \cite{rastegin1} we introduced
such a criterion called "relative error." We have found that
minimizing the relative error is essentially different task from
optimizing other quantities. The asymmetric cloner, which minimizes
the relative error, was us constructed. As Ref. \cite{rastegin1}
shows, the study of the relative error has allowed to complement a
portrait of the state-dependent cloning.

All the above results examine the pure-state cloning. Ref.
\cite{cirac} introduced the single qubit purification procedure that
was used in extending of the input of the optimal cloners
constructed in Refs. \cite{gisin,werner} to mixed states. However,
the described in Ref. \cite{cirac} scenario is not equivalent to the
standard statement of cloning problem. The approximate copying of
mixed states is interesting for various questions. For example, in
some protocols Alice and Bob encode the bits 1 and 0 into two
non-orthogonal pure states \cite{bennett}. In the reality a
communication channel will inevitably suffer from noise that will
have caused the bits to evolve to mixed states. Eve is then anxious
for cloning of two noncommuting mixed states. Ref. \cite{rastegin2}
extends the concept of state-dependent cloning to the case of mixed
states. The upper bound on the global fidelity for mixed-state
cloning has been established. The notion of the angle between two
mixed states \cite{rastegin2} allows to give simple proof for this
upper bound.

In this paper we define the relative error for mixed-state cloning
and related operations in which the ancilla state contains some
{\it a priori} information about the state to be cloned. The lower
bound on the relative error will be obtained. The optimization of a
general unitary transformation is more difficult problem. It is not
considered in the present work.

\protect\section{Preliminary lemmas}

Before definition of the relative error, we shall prove two useful
statements those maintain our approach. In general, the measure of
distinguishability for mixed quantum states is provided by the
fidelity function. The fidelity $F(\chi,\omega)$ between two density
operators $\chi$ and $\omega$ is defined by \cite{jozsa}
\begin{equation}
F(\chi,\omega)
=\max\: \bigl|\langle X|Y \rangle\bigr|^2
\!\ .
\label{fiddef}
\end{equation}
Pure states $|X\rangle$ and $|Y\rangle$ are purifications of
$\chi$ and $\omega$ respectively, that is
$\:\chi={\rm Tr}_{E}\bigl(|X\rangle\langle X|\bigr)\:$ and
$\:\omega={\rm Tr}_{E}\bigl(|Y\rangle\langle Y|\bigr)\:$.
The quantity given by Eq. (\ref{fiddef}) is equivalent to
the Uhlmann's transition probability for mixed states \cite{uhlmann}.
Note that this usage of word "fidelity" is not unique: Refs.
\cite{barnum,uhlmann1} define fidelity to be the square root of the
present quantity. Ref. \cite{rastegin2} parametrized the fidelity by
means of the angle between mixed states, namely:
\begin{align}
F(\chi,\omega) &=
\cos^2\!\Delta(\chi,\omega) \!\ ,
\label{link} \\
\Delta(\chi,\omega)
&=\min
\delta(X,Y) \!\ .
\label{mixdef}
\end{align}
Here $\delta(X,Y)\in[0;\pi/2]$ is angle between vectors $|X\rangle$
and $|Y\rangle$. For any triplet $\{\chi,\omega,\rho\}$ of mixed
states \cite{rastegin2},
\begin{equation}
\Delta(\chi,\omega)\leq
\Delta(\chi,\rho)+\Delta(\omega,\rho)
\!\ .
\label{mixplu}
\end{equation}
This result extends the spherical triangle inequality to the case of
mixed states. The first useful statements gives the upper bound on
the difference between fidelities $F(\chi,\rho)$ and
$F(\omega,\rho)$.

{\bf Lemma 1} {\it For any triplet $\{\chi,\omega,\rho\}$ of
mixed states,}
\begin{equation}
\bigl|F(\chi,\rho) - F(\omega,\rho)\bigr|
\leq \sin\Delta(\chi,\omega)
\!\ .
\label{ineqprob}
\end{equation}

{\bf Proof} Because Eq. (\ref{mixplu}) and standard trigonometric
formula $\:\cos^2\!\alpha-\cos^2\!\beta=
-\sin(\alpha+\beta)\sin(\alpha-\beta)\:$ \cite{handbook},
\begin{equation*}
\cos^2\!\Delta_{\chi\rho} -
\cos^2\!\Delta_{\omega\rho} \leq
\cos^2(\Delta_{\chi\omega}-\Delta_{\omega\rho})
 - \cos^2\!\Delta_{\omega\rho}
=\sin\Delta_{\chi\omega}
\sin(2\Delta_{\omega\rho} - \Delta_{\chi\omega})
\leq \sin\Delta_{\chi\omega}
\!\ .
\end{equation*}
We then get by a parallel argument
\begin{equation*}
\cos^2\!\Delta_{\omega\rho} -
\cos^2\!\Delta_{\chi\rho}
\leq \sin\Delta_{\chi\omega}
\!\ ,
\end{equation*}
and the two last inequalities give Eq. (\ref{ineqprob}).
$\blacksquare$

The second useful statement establishes the upper bound on the
modulus of difference between probability distributions generated by
two mixed states $\chi$ and $\omega$ for any measurement. Let
$\{E_a\}$ be a generalized measurement (POVM). Such a
measurement over the system $S$ in state $\rho$ produces outcome $a$
with probability \cite{busch,holevo}
\begin{equation}
p(a|\rho)={\rm Tr}_{S} (E_a\,\rho) \!\ .
\label{prde}
\end{equation}

{\bf Lemma 2} {\it For arbitrary measurement and any two states
$\chi$ and $\omega$,}
\begin{equation}
\bigl|\,p(a|\chi) - p(a|\omega)\,\bigr|
\leq \sin\Delta_{\chi\omega}
\!\ .
\label{inpr}
\end{equation}

{\bf Proof} Recall that POVM can be realized as an orthogonal
measurement over extended system $ST$ \cite{holevo} (this is insured
by Neumark's theorem). That is,
\begin{equation}
{\rm Tr}_{S} (E_a\,\rho)=
{\rm Tr}_{ST}\!\left\{\Pi_a(\rho\otimes\sigma)\right\}
\!\ ,
\label{prdest}
\end{equation}
where $\{\Pi_a\}$ is an orthogonal measurement. We now choose
purifications $|X\rangle$ of $\chi\otimes\sigma$ and $|Y\rangle$ of
$\omega\otimes\sigma$ so that
$\:F(\chi\otimes\sigma,\omega\otimes\sigma)=\bigl|\langle
X|Y\rangle\bigr|^2\:$.  Because Eq. (\ref{prdest}), we can write
\begin{align}
& p(a|\chi)=
\langle X|\,\Pi_a\otimes{\mathbf{1}}\,|X\rangle
\!\ , \label{chiom} \\
& p(a|\omega)=
\langle Y|\,\Pi_a\otimes{\mathbf{1}}\,|Y\rangle
\!\ .
\label{omchi}
\end{align}
Ref. \cite{rastegin1} proved that for arbitrary projector $\Pi$,
\begin{equation}
\bigl|\,
\langle X| \,\Pi\, |X\rangle -
\langle Y| \,\Pi\, |Y\rangle
\,\bigr| \leq \sin\delta_{XY}
\!\ .
\label{prob3}
\end{equation}
Since the fidelity function is multiplicative \cite{jozsa}, there is
$\:F(\chi\otimes\sigma,\omega\otimes\sigma)=F(\chi,\omega)\:$ and
therefore $\,\Delta_{\chi\omega}=\delta_{XY}\,$. Using Eqs.
(\ref{chiom}), (\ref{omchi}) and (\ref{prob3}), we then obtain
(\ref{inpr}). $\blacksquare$

\protect\section{Statement of the problem}

Let us start with a precise description of the physical operation
that will be considered. A register $A$, having an $d$-dimensional
Hilbert space ${\cal{H}}={\mathbb{C}}{\,}^d$ ($d>1$), is initially
prepared in one state from a set
$\,{\mathfrak{A}}=\{\rho_1,\rho_2\}\,$. The ancilla state
$\Upsilon_j$ from a set
$\,{\mathfrak{S}}=\{\Upsilon_1,\Upsilon_2\}\,$ contains some {\it a
priori} (generally non-full) information about the input state of
register $A$. By the ancilla we will mean a system $BE$ composed of
extra register $B$, that is to receive the clone of $\rho_j$, and
environment $E$. If we include an environment space then any physical
operation may be expressed as a unitary evolution. Thus, the final
state of two registers is described by
\begin{equation}
\widetilde{\rho\,}_j={\rm Tr}_{E}
\Bigl(V(\rho_j\otimes\Upsilon_j)
V^{\dagger}\Bigr) \!\ ,
\label{fistat}
\end{equation}
which is partial trace over environment space. In order to
estimate a quality of cloning we shall compare
$\,\widetilde{\rho\,}_j\,$ with the perfect state
$\,\rho_j\otimes\rho_j\,$ that would be produced by the ideal cloning.
(Note that the cloning is special strong form of
broadcasting \cite{barnum}; the examination of approximate
broadcasting is beyond the scope of the present work.)

We shall now justify the notion of the relative
error for the above physical operation. Lemmas 1 and 2 insure that
the sine of angle between two mixed states gives a reasonable measure
of closeness for ones. We shall now use this measure to justify the
notion of the relative error for discussed operations. For brevity,
let us denote
$\:\Delta_j=\Delta(\widetilde{\rho\,}_j,\rho_j\otimes\rho_j)\:$,
where $j=1,2$. According to Eq. (\ref{inpr}), for any measurement
\begin{equation}
\bigl|\,p(a\,|\,\widetilde{\rho\,}_j) -
p(a\,|\,\rho_j\otimes\rho_j)\,\bigr|
\leq \sin\Delta_j
\!\ .
\label{deviat}
\end{equation}
Thus, size $\,\sin\Delta_j\,$ describes upon the
whole the deviation of the resulting probability distribution from
the probability distribution to which it ought to tend. We define
the absolute error as the sum
$\>\sin\Delta_1+\sin\Delta_2\>$. This definition extends the notion
of the absolute error to the case of mixed states. However, this
criterion loses sight of closeness of states $\rho_1$ and
$\rho_2$. Let us take that we want distinguishing the input state of
register $A$ by measurement made on the output. In order to solve the
problem we compare given output $\widetilde{\rho\,}_j$ to both ideal
outputs $\rho_1\otimes\rho_1$ and $\rho_2\otimes\rho_2$. But if the
ideal outputs are not sufficiently distinguishing then it is
difficult. To express this in quantitative form we should
use some measure of closeness for states $\rho_1\otimes\rho_1$ and
$\rho_2\otimes\rho_2$. By Eq. (\ref{inpr}),
$$
\bigl| p(a\,|\,\rho_1\otimes\rho_1) -
p(a\,|\,\rho_2\otimes\rho_2) \bigr| \leq
\sin\Delta(\rho_1\otimes\rho_1,\rho_2\otimes\rho_2)
$$
for any measurement. So, size
$\>\sin\Delta(\rho_1\otimes\rho_1,\rho_2\otimes\rho_2)\>$ provides
such a measure. The closeness of $\widetilde{\rho\,}_1$ to
$\rho_1\otimes\rho_1$ is measured by
$\sin\Delta_1$, the closeness of $\widetilde{\rho\,}_2$ to
$\rho_2\otimes\rho_2$ is measured by
$\sin\Delta_2$. By analogy with the case of pure states
\cite{rastegin1}, the relative error is defined as follows.

{\bf Definition}
{\it The relative error is}
\begin{equation}
R({\mathfrak{A}}|{\mathfrak{S}})=
\frac{\sin\Delta_1+\sin\Delta_2}
{\sin\Delta(\rho_1\otimes\rho_1,\rho_2\otimes\rho_2)}
\!\ .
\label{redef}
\end{equation}

This definition generalizes the notion of the relative error in two
significances. In the first place, it extends the mentioned notion to
the case of mixed states. In the second place, it takes into account
that the ancilla state can contain {\it a priori} information about
the state to be cloned.  We are interested in lower bound on the
relative error defined above.

\protect\section{Main result}

Let us now formulate the basic result of the present work. The
desired lower bound is established by the following theorem.

{\bf Theorem} {\it Let $\,f=\sqrt{F(\rho_1,\rho_2)}\,$,
$\,\phi=\sqrt{F(\Upsilon_1,\Upsilon_2)}\,$.}

(i) {\it For $\,f\leq\phi\leq1\,$ there holds}
\begin{equation}
R({\mathfrak{A}}|{\mathfrak{S}})\geq
f\phi-f^2\sqrt{1-f^2\phi^2}\big/\sqrt{1-f^4}
\!\ ;
\label{theorem}
\end{equation}

(ii) {\it For $\,0\leq\phi\leq f\,$ there holds
$\,R({\mathfrak{A}}|{\mathfrak{S}})\geq0\,$.}

{\bf Proof of the theorem} (i) At first, using Eq. (\ref{mixplu})
twice, we have
\begin{align}
 & \Delta(\rho_1\otimes\rho_1,\rho_2\otimes\rho_2)\leq
\Delta_1+\Delta_2+
\Delta(\widetilde{\rho\,}_1,\widetilde{\rho\,}_2) \nonumber\\
 & \Delta_1+\Delta_2 \geq
\Delta(\rho_1\otimes\rho_1,\rho_2\otimes\rho_2)
-\Delta(\widetilde{\rho\,}_1,\widetilde{\rho\,}_2)
\label{twice2} \!\ .
\end{align}
By the multiplicativity and the unitary preservation,
\begin{equation*}
F(\rho_1,\rho_2)\, F(\Upsilon_1,\Upsilon_2)
=F\bigl(\rho_1\otimes\Upsilon_1,\rho_2\otimes\Upsilon_2\bigr)
= F\Bigl(V(\rho_1\otimes\Upsilon_1)
V^{\dagger},V(\rho_2\otimes\Upsilon_2)
V^{\dagger}\Bigr) \!\ .
\end{equation*}
Because the fidelity cannot decrease under the operation of
partial trace \cite{barnum},
$\:F(\rho_1,\rho_2)\, F(\Upsilon_1,\Upsilon_2)\leq
F(\widetilde{\rho\,}_1,\widetilde{\rho\,}_2)\:$ and
\begin{equation}
\cos\Delta(\widetilde{\rho\,}_1,\widetilde{\rho\,}_2)
\geq f\phi
\!\ . \label{tilde1}
\end{equation}
By Eq. (\ref{tilde1}), we have
\begin{equation}
-\sin\Delta(\widetilde{\rho\,}_1,\widetilde{\rho\,}_2)
\geq -\sqrt{1-f^2\phi^2}
\label{tilde2}
\end{equation}
According to the angle range of values,
\begin{equation}
\sin\Delta_1+\sin\Delta_2\geq
\sin(\Delta_1+\Delta_2)
\!\ .
\label{epsil}
\end{equation}
By Eqs. (\ref{epsil}) and (\ref{twice2}),
\begin{equation*}
R({\mathfrak{A}}|{\mathfrak{S}})\geq
\cos\Delta(\widetilde{\rho\,}_1,\widetilde{\rho\,}_2)
-\sin\Delta(\widetilde{\rho\,}_1,\widetilde{\rho\,}_2)
\,\cot\Delta(\rho_1\otimes\rho_1,\rho_2\otimes\rho_2)
\!\ .
\end{equation*}
Using Eqs. (\ref{tilde1}) and (\ref{tilde2}), the last inequality can
be rewritten as Eq. (\ref{theorem}). Note that if $\phi<f$ then
Eq. (\ref{theorem}) is also valid. However, it is empty, since the
right-hand side of Eq. (\ref{theorem}) becomes negative.

(ii) Suppose that $\phi$ lies between 0 and $f$. It suffices to
show that there are states $\Upsilon_1$ and $\Upsilon_2\,$, such that
\begin{equation}
\rho_j={\rm Tr}_{E}\Upsilon_j
\!\ .
\label{need}
\end{equation}
Then the equality $\,R({\mathfrak{A}}|{\mathfrak{S}})=0\,$
is clearly valid. We may assume without loss of generality that
states $\Upsilon_1$ and $\Upsilon_2$ are pure. If both purifications
$|Y_1\rangle$ of $\rho_1$ and $|Y_2\rangle$ of $\rho_2$ lie in
$\,\cal{H}\otimes\cal{H}\,$ then \cite{jozsa}
\begin{equation}
\bigl|\langle Y_1|Y_2\rangle\bigr|=
\bigl|{\rm Tr}\,(\sqrt{\rho_1}\,\sqrt{\rho_2}\,V)\bigr| \!\ ,
\label{inneryy}
\end{equation}
where trace is taken over space ${\cal{H}}$.
An element $V$ is placed in the unitary group ${\rm U}(d)$ and
freely variable by choice of purifications \cite{jozsa}. We can use
the freedom in $V$ to make the equality $\:\bigl|\langle
Y_1|Y_2\rangle\bigr|=\phi\:$.  To see this possibility we note that
quantity $\:\bigl|\langle Y_1|Y_2\rangle\bigr|\:$ ranges between 0
and $f$.  Indeed, the maximum is equal to the squared root of
fidelity \cite{jozsa}. The minimal value is zero, because we can take
orthogonal purifications.  Recall that ${\rm U}(d)$ is the connected
group. In fact, an arbitrary unitary matrix can de represented as
$\:W\!DW^{-1}\:$, where $W$ is unitary and $\:D={\rm
diag}\,[\exp(i\theta_1),\ldots,\exp(i\theta_d)]\:$ (this is
provided by the spectral theorem for normal matrices \cite{horn}).
Replacing $\theta_k$ by $t\theta_k\,$, we obtain a continuous path in
${\rm U}(d)$, that connects the given matrix ($t=1$) with the
identity matrix ($t=0$).  Since the right-hand side of Eq.
(\ref{inneryy}) is continuous functional and ${\rm U}(d)$ is
connected, each intermediate value is attained by some element of
${\rm U}(d)$. Thus, there are
states $\,\Upsilon_1=|Y_1\rangle\langle Y_1|\,$ and
$\,\Upsilon_2=|Y_2\rangle\langle Y_2|\,$, such that
$\sqrt{F(\Upsilon_1,\Upsilon_2)}=\phi\,$ and  Eq. (\ref{need})
is too valid. $\blacksquare$

At fixed $f$, the right-hand side of Eq. (\ref{theorem}) is
increasing function of parameter $\phi$. For $\,\phi=f\,$ the lower
bound is equal to zero and the equality
$\,R({\mathfrak{A}}|{\mathfrak{S}})=0\,$ can be reached. For
example, it holds when $\,\Upsilon_j=\rho_j\otimes\sigma\,$, i.e.
the full information about the input state is {\it a priori} provided
in the ancilla. Conversely, in the standard cloning there is
no {\it a priori} information, i.e. $\,\Upsilon_j=\Upsilon\,$ and
$\,\phi=1\,$. Then we get the bound
\begin{equation}
R({\mathfrak{A}})\geq
f-f^2\big/\sqrt{1+f^2}
\!\ .
\label{lbclon}
\end{equation}
In general, the parameter $\phi$ marks the top amount of information
which can beforehand be contained in the ancilla. The larger $\phi$
the less this top amount. If $\,\phi=f\,$ then the full information
of the clone can already be provided in the ancilla state. In the
standard cloning, where $\,\phi=1\,$, any knowledge about the input
state $\rho_j$ is {\it a priori} inaccessible. If the lower bound is
seen as function of $\phi$ then its minimum, reached at $\,\phi=f\,$,
is equal to 0 and its maximum, reached at $\,\phi=1\,$, is equal to
the right-hand side of Eq.  (\ref{lbclon}). On the whole, these
conclusions appear as plausible and contribute to the stronger
no-cloning theorem.

Finally, it should be pointed out that our techniques can be
applied to the $N\to L$ operations. In this case the ancilla is
composed of $M=L-N$ extra registers and the environment. As a
result, we obtain the lower bound

(i) {\it For $\,f^M\leq\phi\leq1\,$ there holds}
\begin{equation}
R({\mathfrak{A}}|{\mathfrak{S}})\geq
f^N\phi-f^L
\sqrt{1-f^{2N}\phi^2}\big/\sqrt{1-f^{2L}}
\!\ ;
\label{bound}
\end{equation}

(ii) {\it For $\,0\leq\phi\leq f^M\,$ there holds
$\,R({\mathfrak{A}}|{\mathfrak{S}})\geq0\,$.}

At fixed $f$, the right-hand side of Eq. (\ref{bound}) is
increasing function of parameter $\phi$. For $\,\phi=f^M\,$ the lower
bound is equal to zero. For example, the equality
$\,R({\mathfrak{A}}|{\mathfrak{S}})=0\,$ holds
when $\,\Upsilon_j=\rho_j^{\otimes M}\!\otimes\sigma\,$ and
the full information is already provided in the ancilla state
$\Upsilon_j$. The right-hand side of Eq. (\ref{bound}) is maximal
for the standard cloning in which $\phi=1$.

\protect\section{Conclusion}

We have established the lower bound on the relative error of
mixed-state cloning and related physical operations, in which the
ancilla state contains some {\it a priori} information about the
input state. In the pure-state cloning, $\,\rho_j=|s_j\rangle\langle
s_j|\,$ for $j=1,2$ and parameter $\,f=\bigl|\langle
s_1|s_2\rangle\bigr|\,$. In this case the lower bound given by
right-hand side of Eq. (\ref{lbclon}) is equivalent to the lower
bound obtained in Ref. \cite{rastegin1} for pure-state $1\to2$
cloning. Similarly, at $\phi=1$ Eq. (\ref{bound}) provides the
extension of the lower bound deduced in Ref. \cite{rastegin1} for
pure-state $N\to L$ cloning. For this extension the modulus of the
inner product should be replaced by the square root of fidelity.

In Ref. \cite{rastegin1} we have constructed the asymmetric cloner,
minimizing the relative error. Does this cloner reach the bound
in the case of mixed states?  The answer is negative. The above
transformation has two properties:
\begin{itemize}
\item[P1]{It acts on $\,\cal{H}\otimes\cal{H}\,$;}
\item[P2]{The initial
state of register $B$ is pure.}
\end{itemize}
It can be shown that two these properties do not allow to reach the
lower bound given by Eq. (\ref{lbclon}) for each pair of kind
$\:{\mathfrak{A}}=\{{\mathbf{1}}/d,|s\rangle\langle s|\}\:$. We
refrain from presenting the proof that is somewhat lengthy. Now we do
not know whether a cloner, which reaches the established lower bound,
exists. Recall that in the case of mixed states the optimal "global"
cloner given in Ref. \cite{brass} does not reach the upper bound on
the global fidelity \cite{rastegin2}. Thus, there is a essential
difference between pure-state and mixed-state cloning.

\end{document}